\documentclass[twocolumn]{aastex701}
\usepackage{hyperref}

\newcommand{\tess}{\textit{TESS}}
\newcommand{\event}{AT~2020afjz}
\newcommand{\vegas}{\texttt{VegasAfterglow}}
\newcommand{\naut}{\texttt{Nautilus}}
\newcommand{\tessellate}{\texttt{TESSELLATE}}
\newcommand{\highlat}{Roxburgh and Ridden-Harper et al. (in prep.)}
\defcitealias{Planck}{Planck Collaboration 2020}
\defcitealias{Astropy2013}{Astropy Collaboration 2013}
\defcitealias{Astropy2018}{2018}
\defcitealias{Astropy2022}{2022}

\usepackage{multirow}

\shorttitle{\event\ (TSS2020a)}
\shortauthors{Ridden-Harper \& Roxburgh et al.}

\begin{document}

\title{\event\ (TSS2020a): The First Fast Extragalactic Transient Discovered by \tess}

\author[orcid=0000-0003-1724-2885]{Ryan Ridden-Harper}\thanks{Co-lead authors; these authors contributed equally to this work.}
\affiliation{School of Physical and Chemical Sciences — Te Kura Matū, University of Canterbury, Private Bag 4800, Christchurch 8140, \\ Aotearoa, New Zealand}
\email[show]{ryanridden@gmail.com}
\correspondingauthor{Ryan Ridden-Harper}

\author[orcid=0009-0001-6992-0898]{Hugh Roxburgh}\thanks{Co-lead authors; these authors contributed equally to this work.}
\affiliation{International Centre for Radio Astronomy Research, Curtin University, Bentley, WA 6102, Australia}
%\affiliation{School of Physical and Chemical Sciences --- Te Kura Mat\={u}, University of Canterbury, Private Bag 4800, Christchurch 8140, New Zealand}
\email{hugh.roxburgh@postgrad.curtin.edu.au}

\author[orcid=0009-0008-4935-069X]{Clarinda Montilla}
\affiliation{School of Physical and Chemical Sciences — Te Kura Matū, University of Canterbury, Private Bag 4800, Christchurch 8140, \\ Aotearoa, New Zealand}
\email{}%{clarinda.montilla@pg.canterbury.ac.nz}  

\author[orcid=0009-0005-0556-3886]{Lancia Hubley}
\affiliation{School of Physical and Chemical Sciences — Te Kura Matū, University of Canterbury, Private Bag 4800, Christchurch 8140, \\ Aotearoa, New Zealand}
\email{}

\author[orcid=0009-0006-7990-0547]{James Freeburn}
\affiliation{Department of Physics and Astronomy, University of North Carolina at Chapel Hill, Chapel Hill, NC 27599-3255, USA}
\email{}

\author[orcid=0009-0008-6765-5171]{Brayden Leicester}
\affiliation{Centre for Astrophysics and Supercomputing, Swinburne University of Technology, John St, Hawthorn, VIC 3122, Australia}
\affiliation{ARC Centre of Excellence for Gravitational Wave Discovery (OzGrav), John St, Hawthorn, VIC 3122, Australia}
\email{}

\author[orcid=0009-0009-9143-1204]{Andrew Moore}
\affiliation{School of Physical and Chemical Sciences — Te Kura Matū, University of Canterbury, Private Bag 4800, Christchurch 8140, \\ Aotearoa, New Zealand}
\email{}

\author[0009-0003-8380-4003]{Zachary G. Lane}
\affiliation{School of Physical and Chemical Sciences — Te Kura Matū, University of Canterbury, Private Bag 4800, Christchurch 8140, \\ Aotearoa, New Zealand}
\affiliation{Space Telescope Science Institute, 3700 San Martin Drive, Baltimore, MD 21218, USA}
\email{}

\author[orcid=0009-0000-0251-2892]{Jaime Luisi}
\affiliation{School of Physical and Chemical Sciences — Te Kura Matū, University of Canterbury, Private Bag 4800, Christchurch 8140, \\ Aotearoa, New Zealand}
\email{}

\author[orcid=0000-0002-2798-2943]{Koji Shukawa}
\affiliation{Space Telescope Science Institute, 3700 San Martin Drive, Baltimore, MD 21218, USA}
\affiliation{William H. Miller III Department of Physics \& Astronomy, Johns Hopkins University,\\ 3400 N Charles St, Baltimore, MD 21218, USA}
\email{}

\author[orcid=0000-0002-4410-5387]{Armin Rest}
\affiliation{Space Telescope Science Institute, 3700 San Martin Drive, Baltimore, MD 21218, USA}
\affiliation{William H. Miller III Department of Physics \& Astronomy, Johns Hopkins University,\\ 3400 N Charles St, Baltimore, MD 21218, USA}
\email{}
% \email{arest@stsci.edu}  

\author[orcid=0000-0001-5703-2108]{Jeff Cooke}
\affiliation{Centre for Astrophysics and Supercomputing, Swinburne University of Technology, John St, Hawthorn, VIC 3122, Australia}
\affiliation{ARC Centre of Excellence for Gravitational Wave Discovery (OzGrav), John St, Hawthorn, VIC 3122, Australia}
\email{}

\author[orcid=0000-0003-3257-4490]{Michele T. Bannister}
\affiliation{School of Physical and Chemical Sciences — Te Kura Matū, University of Canterbury, Private Bag 4800, Christchurch 8140, \\ Aotearoa, New Zealand}
\email{michele.bannister@canterbury.ac.nz}

\author[orcid=0009-0002-8965-5574]{Lilly Fox}
\affiliation{School of Physical and Chemical Sciences — Te Kura Matū, University of Canterbury, Private Bag 4800, Christchurch 8140, \\ Aotearoa, New Zealand}
\email{}

\author[orcid=0009-0007-9868-0014]{Tait Keller}
\affiliation{School of Physical and Chemical Sciences — Te Kura Matū, University of Canterbury, Private Bag 4800, Christchurch 8140, \\ Aotearoa, New Zealand}
\email{}

\author[orcid=0000-0001-5233-6989]{Qinan Wang}
\affiliation{Department of Physics and Kavli Institute for Astrophysics and Space Research, Massachusetts Institute of Technology, \\ Cambridge, MA 02139, USA}
\email{}

\begin{abstract}

We report the discovery of \event\ (TSS2020a): the first hour-scale extragalactic transient discovered in optical wavelengths whose complete evolution --- from explosion onset to decay --- is temporally resolved, and the first such transient discovered by \tess. \event\ was identified as a $>$$10\sigma$ detection in the pilot HiLaTS program run within the TESSELLATE Sky Survey, which blindly searches for transient phenomena in \tess\ data with the \tessellate\ pipeline. Through cross-matching with legacy imaging, we associate it with DES~J042144.37$-$383311.3, a member of an interacting galaxy pair at $z_{\rm phot}=0.67^{+0.07}_{-0.10}$. While \event\ is similar in duration and brightness to GRB afterglows, it exhibits a slow rise time of $\sim1$~hr and lasts for only 2.4~hr above the half-max brightness; modeling the \tess\ light curve with \vegas\ finds that it is best described as either an on-axis ``dirty-fireball" or off-axis orphan afterglow. Each of these rare classifications hinge upon a non-detection at gamma-ray energies, but as \textit{Fermi}-GBM was Earth-occulted at the time of explosion, \event's gamma-quiet nature cannot be definitively confirmed. Regardless, \event\ demonstrates \tess's power to discover fast extragalactic transients, and heralds a new population awaiting discovery with \tessellate.

\end{abstract}



\section{Introduction} 

While much is known about optical transient classes that evolve over timescales of days to weeks, comparatively little is known about events that evolve more rapidly. 
To date, few large-scale transient surveys have been conducted at high cadence with rapid systematic revisits to the same regions of sky. As a result, the fast optical transient phase space remains relatively unexplored \citep{Andreoni2020}, and few classes of astrophysical transients are known to produce optical emission that evolves on hour-long timescales \citep{Villar2017}. 

Luminous optical transients with hour-scale durations must be powered by relativistic processes such as shocks \citep{Nakar2012} and jets \citep{Bromberg2011}, and therefore provide unique probes of extreme physics. Among these, GRB afterglows represent the primary established class \citep{Piran2004, Kann2010, Kann2011}. However, they are typically discovered only after high-energy satellite triggers and targeted follow-up observations, rather than through untargeted optical surveys. 
Such searches for hour-timescale optical transients remain rare, and consequently the rates and demographics of such events are poorly constrained.

The observed bimodal distribution of GRB durations \citep[$T_{90}$,][]{Kouveliotou93}, together with their host-galaxy environments and localizations, has led to the identification of two distinct progenitor channels. Long GRBs ($T_{90}>2\,$s) are generally associated with the core collapse of massive stars \citep[collapsars,][]{Woosley93,MacFadyen1999,Woosley2006}, whereas short GRBs ($T_{90}<2\,$s) are thought to arise from the merger of compact objects \citep{Eichler89}. Although their progenitors differ, both classes are believed to produce highly collimated relativistic jets with small opening angles and large Lorentz factors. As these jets propagate into the surrounding circumburst medium (CBM), they drive shocks that accelerate particles and produce broadband synchrotron emission observed as an afterglow \citep{Meszaros97}. 

Since GRBs and their afterglows are the products of complex and high-energy interactions, a number of subclasses naturally emerge. While the prompt $\gamma$-ray emission is visible only to observers within the jet opening angle, in theory the afterglow can remain detectable for off-axis observers, producing so-called orphan afterglows \citep{Rhoads1997}. Alternatively, if the GRB jet is choked with excess baryons, then the $\gamma$-rays will be suppressed, and the Lorentz factor will be severely diminished, creating a slow-evolving ``dirty fireball'' \citep{Dermer2000}. Theoretical work suggests orphan afterglows and dirty fireballs should be at least as numerous as classical on-axis GRBs: the small jet opening angles inferred for GRBs imply that on-axis bursts represent only $\sim0.1$–$0.2\%$ of all GRB jets \citep{Rhoads1997,Huang2002}. Observationally the reverse holds: $\gamma$-ray monitors have cataloged $\gtrsim4000$ on-axis GRBs (Fermi-GBM; plus $\sim1700$ from Swift), yet convincing detections of unconventional afterglows remain exceedingly rare, with only a handful of candidates from untargeted optical surveys \citep[e.g.][]{Cenko2013,Ho2020,Perley2025}.

It remains unclear whether this discrepancy reflects an incomplete physical picture or an observational bias. For decades, missions such as \textit{Fermi}-GBM \citep{Meegan2009} and \textit{Swift} \citep{Gehrels2004} have provided continuous monitoring of the high-energy sky, enabling the routine detection of prompt $\gamma$-ray emission. In contrast, optical facilities capable of systematically monitoring large areas of the sky at cadences comparable to the evolution of GRB afterglows are only now becoming more prevalent. One such instrument is the \textit{Transiting Exoplanet Survey Satellite} \citep[\tess, ][]{Ricker2014}, whose rapid observation cadence, continuous monitoring, and wide field of view (FoV) makes it a superb instrument to hunt peculiar afterglows.

Since its launch in July 2018, \tess\ has successfully documented a number of GRB afterglows \citep[e.g.][]{Smith2021,Roxburgh2024,Jayaraman2024,Perley2025}, as well as other extragalactic transients \citep[e.g.][]{Wang2023,Fausnaugh2023,Wang2024,Lane2026}. \tess\ has also provided valuable constraints on exotic fast transients like optical counterparts to Fast Radio Busts \citep[FRBs; ][]{Tingay2019} and flares from Luminous Fast Blue Optical Transients \citep[LFBOTs; ][]{Jayaraman2026}. Thanks to its near-continuous observation strategy and temporal cadence --- initially 30 minutes, then 10 minutes from July 2020, and now 200 seconds since September 2022 --- it has provided unparalleled insights into the evolutions of such events. However, its poor spatial resolution ($21''$ per pixel) and challenging scattered-light background have largely limited its use to follow-up observations rather than untargeted transient identification. Consequently, \textit{TESS} high-cadence data have supplemented, rather than driven, transient discoveries.

The \texttt{TESSELLATE} pipeline aims to improve the capabilities of \tess\ by conducting an untargeted search of all archival \tess\ data for unknown transient signals \defcitealias{TESSELLATE}{H. Roxburgh and R. Ridden-Harper et al. 2025}\citepalias{TESSELLATE}. Built on the \texttt{TESSreduce} reduction pipeline --- which comprehensively models and subtracts a spatially and temporally varying background to enable maximally sensitive difference imaging \citep{tessreduce} --- \texttt{TESSELLATE} is optimized for processing \tess\ full-frame images (FFIs) in bulk and combing the differenced images for variability across all timescales. All detected events are cataloged as part of the \texttt{TESSELLATE} Sky Survey (TSS), whose primary goals include discovering previously unknown transients, documenting stellar variability at the fastest time scales, and detecting new exoplanets. 

The pilot survey of the TSS covers all of \tess's Year 3 and 4 data, spanning Sectors 27-55 and observing the entire sky. A key program of the pilot survey is the High galactic Latitude Transient Search (HiLaTS) program, whose focus is on detecting rapid extragalactic transients and constraining the volumetric rates of various transient classes, such as GRB afterglows, and fast blue optical transients \citep[FBOTs; ][]{Drout2014,Rest2018,Inserra2019, Ho2023}. This program filters events recorded by \texttt{TESSELLATE} based on a number of measured criteria, including light curve significance, point-spread-function (PSF) quality, event duration, and positional accuracy. Further details of the program, including the significant sensitivity and localization improvements made to \texttt{TESSELLATE} since its inception, will be discussed in \highlat.

In this paper, we present the first untargeted discovery of a rapid extragalactic optical transient by \tess, identified through the TSS-HiLaTS program. In Section~\ref{sec:data}, we describe the data processing used to characterize and localize the event. In Section~\ref{sec:galaxies} we examine the properties of its potential host galaxies. In Section~\ref{sec:analysis}, we model the light curve and discuss the nature of its origin, comparing with the properties of various transient classes. Finally, in Section~\ref{sec:dicsussion} we forecast the future prospects of the TSS and summarize our findings. Throughout this work, we assume standard Planck cosmology \citepalias{Planck} and present all magnitudes in the AB system unless stated otherwise.

\section{Data}\label{sec:data} 

\subsection{Discovery With \tess}

\begin{figure*}[h]
    \includegraphics[width=1\textwidth]{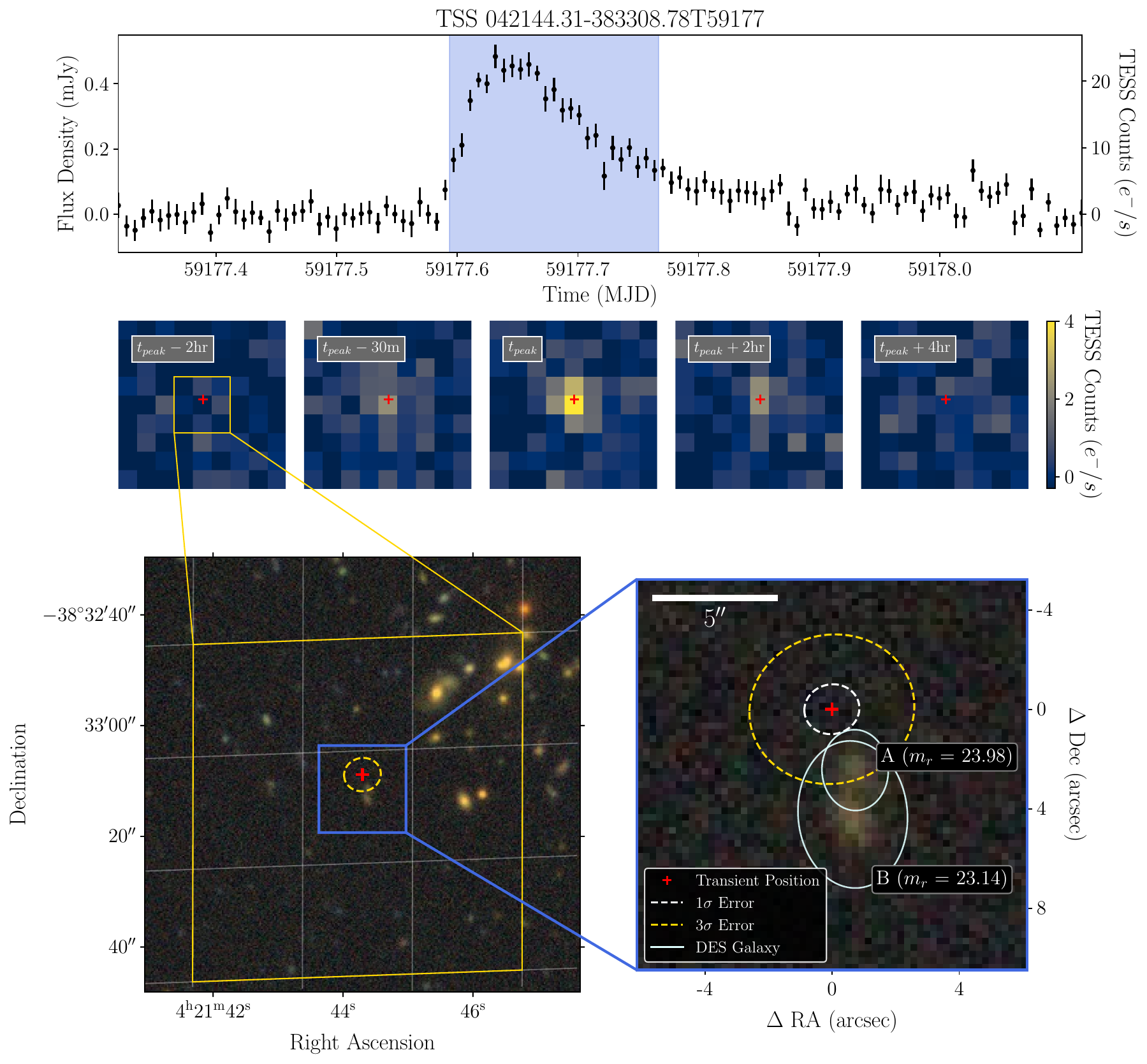}
    \caption{The extragalactic optical transient TSS 042144.31-383308.66T59177. \textbf{Top Row:} Flux-calibrated light curve extracted through PSF photometry. The shaded blue region indicates the time span during which the event was detected in \texttt{TESSELLATE}. \textbf{Middle Row:} \tess\ images of the region at various epochs before, during, and after the event. The red plus indicates the event's PSF fit centroid returned by \texttt{TESSELLATE}. \textbf{Bottom Left:} Legacy DR10 DECam grz false color imaging of the region around the event's localization. The faint white grid lines indicate the boundaries of \tess pixels, and the red plus and yellow ellipse indicate the event's centroid and 3$\sigma$ error respectively. \textbf{Bottom Right:} Zoom in of the localization region, revealing two faint galaxies cataloged in DES, which are bounded by their listed Kron ellipses in pale blue.}
    \label{fig:main}
\end{figure*}

On November 24th 2020, TSS 042144.31-383308.66T59177 (hereafter \event) was observed by \tess\ during Sector 32, beginning its rise at 14:09:32.122 UTC, remaining visible ($>3\sigma$) for $\sim$10~hours before fading into the background. Its brightness peaked $\sim$60 minutes into its duration, remaining above half-max brightness for 2.4~hours; given the 10 minute cadence of the data, \tess\ therefore temporally resolved the rise and evolution to high precision. The light curve of \event\ is displayed in the top panel of \autoref{fig:main}, while five differenced \tess\ images at various times across the event's duration are shown in the middle panels.

\event\ was identified in the pilot TSS-HiLaTS program, after successfully passing several event-property filters. These filters are constructed to restrict the selection of cosmic rays, flare stars, and other instrumental or undesired astrophysical signals. With a peak light curve signal-to-noise ratio (SNR) of 18.6, \event\ was readily identified through manual vetting of candidates. 

Flux calibration in \texttt{TESSELLATE} uses the RP magnitudes of cataloged \textit{Gaia} sources \citep[\textit{Gaia;}][]{GaiaCollaboration2016,GaiaCollaboration2023} on a cut-by-cut basis, where each cut is $256\times256$ pixels in size. The \tess\ and \textit{Gaia} RP bandpasses share nearly identical wavelength ranges, though the latter lacks approximately 50~nm at the blue end. This difference only marginally affects the quality of \tess's zero-point determination, and is significantly simpler and cheaper than the standard multi-band polynomial filter matching traditionally employed by \texttt{TESSreduce}. Further details will be described in \highlat. For this cut, 729 cataloged stars brighter than a \textit{Gaia} RP AB magnitude of 15.5 were matched to the chosen background-subtracted \tess\ reference image, from which its zero-point was determined to be $20.626 \pm0.001$ mag\footnote{This value differs from the conventional $20.44\pm0.05$ zeropoint in Vega magnitudes that is used in other studies. We will explore the AB zeropoint and instrument calibration in detail in upcoming work.}. This implies \event's brightness peaks at a magnitude of $17.19\pm0.08$. A summary of its measured properties are displayed in \autoref{tab:properties}.

\begin{deluxetable}{lc}[t]
\caption{Properties of \event.}
\label{tab:properties}
\tablewidth{\columnwidth}
\tablehead{\colhead{Quantity} & \colhead{Value}}
\startdata
R.A. (J2000) & $04^{\text{h}}21^{\text{m}}44.31^{\text{s}} \pm\ 1.1''$ \\
Decl. (J2000) & $-38^{\circ}33'08.66'' \pm\ 1.0''$\\
\textit{TESS} Sector, Cam, CCD & 32,3,2 \\
Peak MJD & 59177.632  \\
Peak UTC & 2020 Nov 24 15:10:04  \\
Duration above half-max ($t_{1/2}$) & 2.4 hrs \\
PSF-like & 0.99 \\ 
Peak \textit{TESS} Counts & $18.4 \pm 1.5$ \\
Peak SNR & 18.6 \\
Peak Flux Density & $0.48 \pm 0.03$ mJy\\
Peak Magnitude (\tess) & $17.19 \pm 0.08$ \\[1mm] \hline 
Likely Host & DES J042144.37-383311.3\\
$z_{\rm phot}$ & $0.67^{+0.07}_{-0.10}$\\
Absolute Magnitude (\tess) & $-25.7^{+0.5}_{-0.28}$
\\[1mm]
\hline\\[-4mm]
\enddata
\end{deluxetable}

\subsection{Host Association $\&$ Extragalactic Origin}

\texttt{TESSELLATE}'s localization procedure, which incorporates PSF fitting with robust positional errors, localized \event\ to $\mathrm{R.A.}=04^{\mathrm{h}}21^{\mathrm{m}}44.31^{\mathrm{s}}$, $\mathrm{decl.} = -38^{\circ}33'08.66''$, with $1\sigma$ uncertainties of $1.1''$ and $1.0''$, respectively. We verified the reliability of the localization by searching for nearby bright flaring events on the detector; one such rapid outburst was identified just 14 pixels ($\sim$$5'$) away, and was precisely localized by \texttt{TESSELLATE} to a red M-dwarf flare star (see ~\autoref{fig:complex} in Appendix). This verifies that the local World Coordinate System is accurately calibrated.

\event's localization lies near a faint, extended source visible in imaging from the DESI Legacy Imaging Surveys \citep{Dey2019}, as shown in the lower panels of \autoref{fig:main}. While the Legacy Surveys DR9 and DR10 consider this source a single object, an independent reduction from the DECam Local Volume Exploration Survey \citep[DELVE;][]{Drlica-Wagner2021} resolves it into two discrete components, DES J042144.37-383311.3 and DES J042144.37-383313.0 (hereafter A and B). 

No other significant feature is present in the Legacy Surveys' imaging within the area bounded by the central \tess\ pixel. The nearest alternative object in the DES DR2 catalog --- which reaches a depth of $m_r\approx24.7$ --- is separated from \event\ by $>8\sigma$. At such a high galactic latitude ($b=-45\degr$), the existence of an incredibly faint Galactic progenitor is very unlikely; unWISE imaging of the nearby field \citep[unWISE;][]{Schlafly2019} reveals only a single component coincident with the extended source, likely ruling out the existence of a faint dwarf star closer to the event centroid. As such, \event\ appears to be of extragalactic origin, either arising from candidate host A or a fainter host unseen by DECam.

The intrinsic properties of \event\ support this conclusion. Given its peak magnitude of $\approx17$, a quiescent Galactic counterpart such as a cataclysmic variable (CV) would need to brighten by at least 8 magnitudes. Dwarf novae typically exhibit outburst amplitudes of $\sim$2--5~mag \citep{OtulakowskaHypka2016}, and also evolve much more slowly; \event\ remains above half-maximum for only $2.4$\,hr compared to typical dwarf nova rise times of $\sim1$\,day and decay times spanning days to weeks. Furthermore, no recurrent outbursts are seen at \event's position across the full \tess\ baseline or in subsequent sectors, and no spatially coincident transient has ever been reported to the Transient Name Server. We therefore rule out a Galactic CV progenitor.

As component A's center is separated by just $2.6^{+1.2}_{-0.7}$ arcseconds from \event's PSF-fit centroid --- falling well within the 3$\sigma$ uncertainty region --- we thus consider it the primary candidate host galaxy of the event. We assess the probability of their chance alignment following \citet{Bloom2002}. The surface density of galaxies equally bright or brighter than the host's apparent magnitude ($r=23.98$) is $\sigma\simeq6\times10^{4}\,\mathrm{deg^{-2}}$; at the measured angular offset, this yields a chance-coincidence probability $P_{\rm cc}=0.10^{+0.09}_{-0.05}$. While it thus seems likely that A is the probable host, it is not conclusive evidence. Regardless, the evidence suggests that \event\ has an extragalactic origin, whether the host is component A or a fainter unseen galaxy.

\subsection{Archival and Multi-wavelength Crossmatching}

Following its discovery, we searched archival databases for transient notices coincident with the time and localization of \event, finding nothing registered on the Transient Name Server \citep[TNS;][]{TNS, Gal-Yam2021} within 10 arcseconds of its position. As such we logged the event with TNS, receiving the designation \event.

We then searched for contemporaneous sky coverage from other instruments, noting that at a declination below $-30^{\circ}$, we are limited to follow-up resources that cover the southern hemisphere. As such, we checked for coverage with \textit{Fermi}-GBM \citep{Atwood2009,Meegan2009}, \textit{Swift} \citep{Gehrels2004,Barthelmy2005}, ATLAS \citep{Tonry2018} and other all-sky monitors. No coverage was available at radio wavelengths from the VAST survey \citep{Murphy2021} on ASKAP \citep{Hotan2021}, as the survey's extragalactic program had not yet commenced, and LIGO \citep{LIGO} was in downtime between its O3 and O4 observing runs.

The \textit{Fermi}-GBM field of view regularly observes the coordinates of \event, with $\sim$35 min gaps due to Earth occultation \citep{Meegan2009}. Unfortunately, \event's estimated explosion time ($t_0$) fell inside one of these gaps, and thus \textit{Fermi} is unable to constrain potential prompt gamma-ray emission associated with this event. Away from the event time, \textit{Fermi}-GBM has detected 176 bursts that spatially overlap with the position of \event, though their localization uncertainties are large and consistent with random chance alignment. 

No trigger was detected by the Konus-Wind \citep{Aptekar1995} and the wider InterPlanetary Network GBM \citep{Hurley2013} which continuously monitor the entire sky. 
The non-detection places the upper bound at the Konus-Wind triggered-mode sensitivity of ${\sim}10^{-6}\,\mathrm{erg\,cm^{-2}\,s^{-1}}$ \cite[$20\,\mathrm{keV}$--$10\,\mathrm{MeV}$; ][]{Tsvetkova2021}.

ATLAS has monitored the field for $\sim10$ years, but has not detected any transient at \event's location. It observed the field 8.1 days before \event\ and 1.76 days after, but missed the short window of observability in between. Similarly, no \textit{Swift} data was available before, during, or after this time. 
Our search therefore suggests that \tess\ was the sole instrument to observe \event; as such, all interpretation we present regarding its nature and origin is based on the \tess\ light curve and localization region alone.

\section{Host properties}\label{sec:galaxies}

\begin{figure}
    \includegraphics[width=0.48\textwidth]{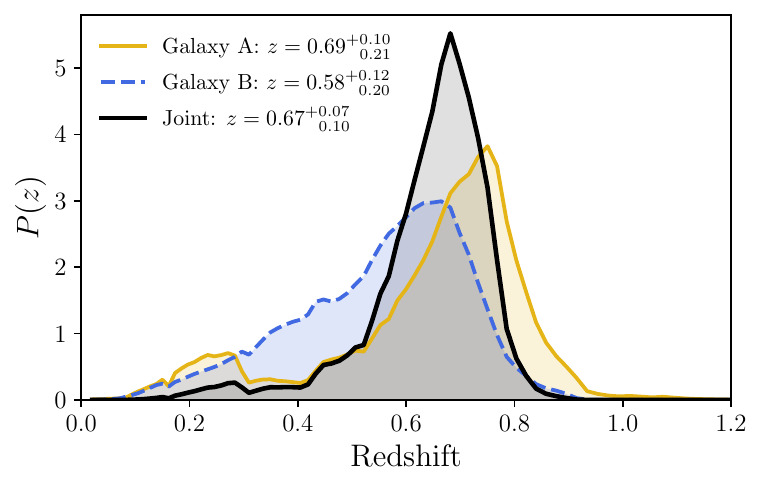}
    \caption{Photometric-redshift posteriors $P(z)$ for the two candidate host galaxies A (gold solid) and B (blue dashed) for \event, alongside their joint fit (black solid). All fits use \texttt{EAZY} template fitting with an $r$-band apparent-magnitude prior. Curves are normalized to unit area and shaded below; the legends show the median and $16$--$84$th-percentile credible interval.}
    \label{fig:photz}
\end{figure}

Given \event's relatively low-chance coincidence with component A of the blended source seen in Legacy DR10 DECam imaging, we consider this galaxy the nominal host for the remainder of this work. 
Although no spectroscopic redshift is available, two independent DESI Legacy Imaging Survey catalogs estimate the photometric redshift of the blended source to be $0.75\leq z_{\text{phot}}\leq0.9$, derived using machine-learning and template-fitting methods applied to the blended DR9 and DR10 photometry \citep{Zhou23,Li24}. Rather than using these estimates, we instead compile de-blended photometric observations from DECam and WISE imaging (\autoref{tab:galdetails}) and compute $z_{\text{phot}}$ separately for each DELVE component using the Easy and Accurate Z$\rm _{phot}$ from Yale package \citep[\texttt{EAZY; }][]{Brammer2008}, incorporating apparent magnitude priors based on DECam $r$-band measurements. We find relatively strong similarity in the posteriors for both components, and thus also compute a combined $z_{\text{phot}}$ that treats the components as separate but interacting at a common redshift. This approach is distinct from the literature estimates which assume a single blended source. The resulting posterior distributions $P(z)$ are shown in \autoref{fig:photz}. We find that the joint fit offers the most constrained redshift estimate, and henceforth adopt a nominal $z_{\text{phot}}=0.67^{+0.07}_{-0.10}$. In Section~\ref{sec:analysis}, we use this joint photometric redshift posterior distribution as a model prior.

\begin{deluxetable}{lcc}[t]
\footnotesize{\tablecaption{Photometric properties of the candidate host galaxies for \event. For the joint fit we assume the galaxies are an interacting pair and share a common redshift. We report $2\sigma$ upper bounds where no source was detected. \label{tab:galdetails}}}
\tablewidth{0pt}
\tabletypesize{\footnotesize}
\tablehead{
\colhead{Property} & \colhead{A (likely host)} & \colhead{B}
}
\startdata
DES ID     & J042144.37$-$383311.30 & J042144.37$-$383313.0 \\
R.A.                      & 04:21:44.38 & 04:21:44.37 \\
Decl.                     & $-$38:33:11.3 & $-$38:33:13.1 \\
Offset & 2.61\arcsec & 4.47\arcsec \\
\hline
$g$                               & $25.02\pm0.19$ & $24.50\pm0.21$ \\
$r$                               & $23.98\pm0.10$ & $23.14\pm0.08$ \\
$i$                               & $23.21\pm0.09$ & $22.46\pm0.08$ \\
$z$                               & $23.12\pm0.15$ & $22.16\pm0.11$ \\
$y$                               & $23.08\pm0.45$ & $22.59\pm0.51$ \\
$W1$                              & $22.48\pm0.68$ & $21.14\pm0.20$ \\
$W2$                              & $>21.3$         & $21.82\pm0.84$ \\
\hline \\[-3mm]
\multirow{2}{*}{$z_{\rm phot}$} & $0.69^{+0.10}_{-0.21}$ & $0.58^{+0.12}_{-0.20}$ \\[1mm]
 & \multicolumn{2}{c}{$0.67^{+0.07}_{-0.10}$ (Joint)} \\%[-3mm]
 \hline \\[-3mm]
\multicolumn{3}{c}{STELLAR POPULATION (\textsc{prospector}, $z=0.67$)} \\[1mm]
$\log_{10}(M_\star/M_\odot)$                  & $9.72^{+0.09}_{-0.10}$ & $10.13^{+0.18}_{-0.10}$ \\
$\log_{10}({\rm SFR}/M_\odot\,{\rm yr^{-1}})$ & $0.0^{+0.5}_{-0.6}$    & $0.3^{+0.5}_{-0.7}$ \\
$\log_{10}({\rm sSFR}/{\rm yr^{-1}})$         & $-9.7^{+0.5}_{-0.5}$   & $-9.9^{+0.5}_{-0.7}$ \\
$A_V$ (mag)                                   & $0.9^{+0.6}_{-0.6}$    & $1.1^{+0.6}_{-0.7}$ \\
Classification & star-forming & star-forming\\[1mm]
\hline\\[-3.5mm]
\enddata
\end{deluxetable}

\begin{figure}[t]
    \includegraphics[width=0.48\textwidth]{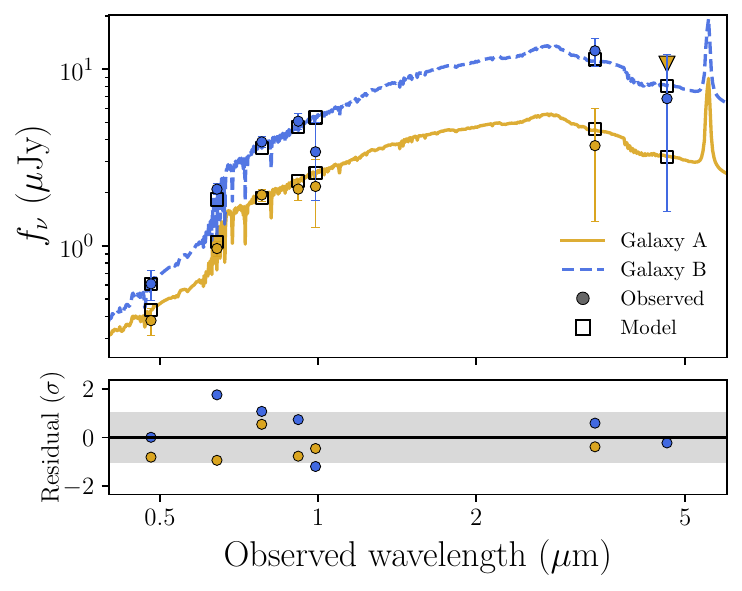}
    \caption{Best-fitting \texttt{Prospector} SEDs of the two candidate hosts, Galaxy~A (gold, solid) and Galaxy~B (blue, dashed), at $z=0.67$. Filled circles are the dereddened photometry (DES $grizy$ + unWISE $W1/W2$), open squares the model fluxes, and triangles $2\sigma$ upper limits. Curves show the posterior-median FSPS delayed-$\tau$ model. The lower panel gives residuals in units of $\sigma$ (shaded
$\pm1\sigma$).}
    \label{fig:prospector}
\end{figure}

To estimate the stellar-population properties of galaxies A and B, we employ the spectral energy distribution (SED) fitting package \texttt{Prospector} \citep{Johnson2021}. While the available SED information is limited, \texttt{Prospector} is capable of providing approximate properties from photometry alone. We correct for Milky-Way extinction \citep{Schlafly2011} and fit the DES $grizy$ and deblended unWISE $W1/W2$ photometry with \texttt{FSPS} stellar populations \citep{Conroy2009,Conroy2010}, fixing the redshift to the median joint photometric value $z=0.67$ and adopting a parametric delayed-$\tau$ star-formation history, SFR$(t)\propto t\,e^{-t/\tau}$. The free parameters are the total stellar mass, stellar metallicity, dust attenuation, the population age $t_{\rm age}$, and the $e$-folding time $\tau$, sampled with \texttt{emcee} \citep{Foreman-Mackey2013}. 

Our best-fit SEDs for both galaxies are shown in \autoref{fig:prospector}, with the marginalized posteriors in \autoref{fig:prospect_corner}. Given the small number of broadband points, only the stellar mass is well-constrained ($\log M_\star/M_\odot\simeq9.7$ and $10.1$ for A and B, respectively); the metallicity, dust, and age are effectively prior-dominated, so we treat the derived star-formation rates as order-of-magnitude estimates. The computed star-formation rates and specific star-formation rates indicate that both galaxies are star-forming rather than quiescent. Their consistent redshifts and small projected separation are suggestive of an interacting pair.

It is worth noting that at a redshift of 0.67, component A is separated from the PSF-fit centroid of \event\ by $18.9^{+8.6}_{-5.1}$ kpc. This offset complicates our host association, as it places \event\ far from the galactic core. Without deeper, targeted observations of the field, we cannot rule out the existence of a fainter dwarf galaxy lying closer to the transient position that is unseen in DECam imaging. Since $z$ is needed for modeling, we move forward with this association, and discuss the implications of such an extreme host offset for the nature of the progenitor in the following section.

\section{Analysis}\label{sec:analysis}

\subsection{Transient Classification}

With only the \tess\ light curve, conclusively classifying \event\ is challenging. 
Despite this limitation, it is clear that \event\ is powered by a relativistic engine: it increased its brightness by at least 2 magnitudes within an hour, exceeding the threshold for Arnett's rule for radioactive powered transients \citep{Arnett1982}. 

\begin{figure}[t]
    \includegraphics[width=0.48\textwidth]{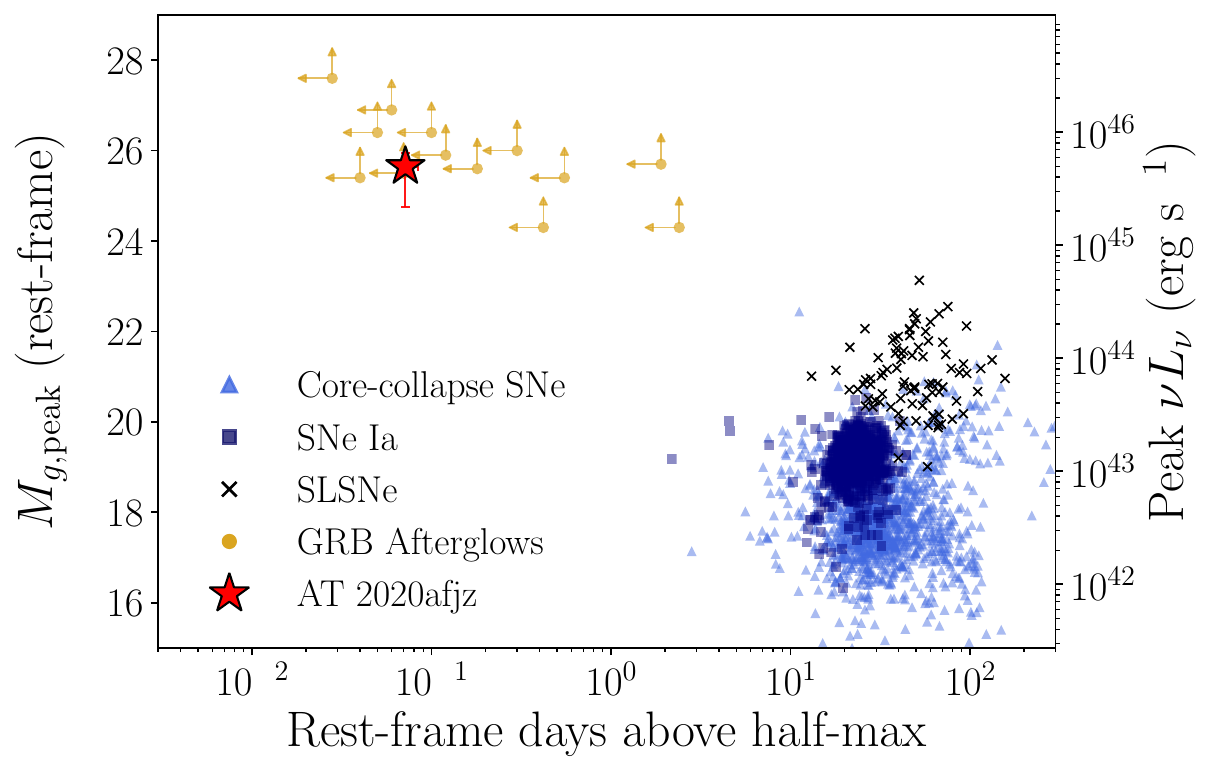}
    \caption{Peak absolute magnitude ($M_{g,\mathrm{peak}}$, rest-frame) versus rest-frame duration above half-maximum, following \citet{Ho2022}. A range of extragalactic transients are shown, where \event\ falls within the population of relativistic transients marked by the GRB afterglows.}
    \label{fig:phasespace}
\end{figure}

In \autoref{fig:phasespace}, we compare \event's evolution rate and approximate rest-frame brightness with other optical extragalactic transients, finding its broad properties closely align with those of GRB afterglows \citep{Ho2022}. Such diagnostics have proven effective at separating afterglows from other classes of fast transients, though rarer classes not included in this comparison cannot be fully excluded on this basis alone. With only the \tess\ light curve, it is unproductive to speculate further about potential progenitor mechanisms; we therefore proceed under the assumption that \event\ is a GRB afterglow.

\begin{figure}
    \includegraphics[width=0.47\textwidth]{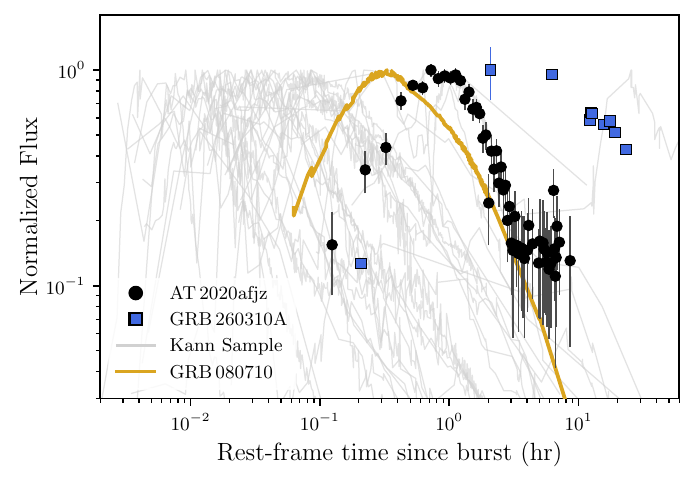}
    \caption{Peak-normalized optical light curves versus rest-frame time since burst for \event\ (black), GRB\,260310A (blue squares; \citealt{Yang2026}), and the \citet{Kann2010} sample of GRB afterglows with measured rises (gray), with GRB\,080710 --- the afterglow whose rest-frame rise timescale most closely matches our event --- highlighted in gold. \event's rise morphology and peak timing track the slow-rising, off-axis-interpreted afterglows 080710 and 260310A rather than the bulk of the sample, supporting a relativistic-afterglow origin.}
    \label{fig:comparison}
\end{figure}

As such, in \autoref{fig:comparison} we compare its light curve to the well-sampled composite $R$-band GRB afterglow light curves of \citet[][hereafter Kann]{Kann2010}, restricting the sample from 176 to 31 afterglows with measured redshifts and resolved rises. \event\ exhibits an extreme light curve morphology, with a notably slow rise and long duration; within the Kann sample, only GRB~080710 appears similar, though \event\ is longer lived still. We also highlight the light curve of GRB~260310A, an unusually long afterglow that preceded the bright Type Ic supernova SN~2026fgk \citep{Yang2026,OConnor2026}. The time-resolved evolution of \event\ lies between these two events, suggesting they may share similar progenitor systems.

Both GRB~080710 and GRB~260310A possess features that suggest they were produced by atypical afterglows. Analysis by \citet{Kruhler2009} concludes that GRB~080710 likely was produced by either an off-axis GRB, or an on-axis jet in its pre-deceleration phase with $\Gamma_0<100$, meeting the criteria for a dirty fireball. Conversely, for GRB~260310A \citet{Yang2026} surmise that the late afterglow onset, weak prompt emission, and hard peak energy support an off-axis GRB progenitor. Given the interpretations for these afterglows, we allow for both the off-axis and dirty fireball scenarios when modeling \event\ in \autoref{sec:modeling}.

\subsection{Modeling} \label{sec:modeling}

While we are limited to only \tess\ data, the high SNR and superb time resolution of \event's light curve provide a solid basis for modeling within a GRB afterglow framework, for which we use the \vegas\ package \citep{Wang2026}. We approach this modeling cautiously, as complex, physically-driven models are often degenerate, with inconsistent parameter retrieval. However, through the analysis of other GRB afterglows observed by \tess, we have developed a robust fitting method that reliably recovers simulated input parameters. While we outline the procedure in this section, it will be discussed more thoroughly in Montilla et al. (in prep.).

The \vegas\ package features a range of parameters that describe the macroscopic and microscopic physics at play during GRB afterglows. Unlike other models, \vegas\ incorporates the initial bulk Lorentz factor of the jet, $\Gamma_0$, as a direct model parameter, and is capable of modeling reverse shocks. The framework also supports a variety of jet geometries, as well as flexible circumburst medium density profiles. This makes it a versatile model, well-suited for the modeling of early time observations and peculiar afterglows. For forward shock models, \vegas\ has eight free parameters: the initial bulk Lorentz factor $\Gamma_0$, which sets the energetics and deceleration onset; the isotropic-equivalent kinetic energy $E_{\mathrm{K,iso}}$; the circumburst density, as a constant interstellar medium (ISM) $n$ or wind $A_\star$; the jet core angle $\theta_{\mathrm{c}}$ and viewing angle $\theta_{\mathrm{obs}}$; and the shock microphysics --- the electron index $p$ and the electron and magnetic energy fractions $\epsilon_e$ and $\epsilon_B^{}$. We also augment \vegas\ with both a flexible explosion time parameter $t_0$ and a flexible redshift, incorporating the joint \texttt{EAZY} $z_{\rm phot}$ distribution as the latter's prior. 

Our fits are performed with the \naut\ package \citep{nautilus}, an importance nested-sampling algorithm that uses deep learning to efficiently sample complex, multi-modal posteriors while computing the Bayesian evidence. Unlike traditional MCMC or standard nested-sampling implementations, \naut\ trains neural networks to model the iso-likelihood boundaries and draws samples by importance sampling, accelerating convergence and enabling robust model comparison via Bayes factors. 

As discussed in the previous section, if \event\ is indeed a GRB afterglow, then --- given its light curve morphology --- it is likely an atypical subclass. We therefore explore several model configurations with \vegas, leveraging top-hat, Gaussian, and power-law jet geometries. All models are run with only forward shocks\footnote{Test fits to \event\ with reverse shocks enabled were dominated by the forward shock.}, and no lateral spreading. 

We take particular care in defining priors for degenerate parameters; during testing, the ISM density $n$ was readily driven to unrealistically high values (e.g. $n>10^{5-6}\,$cm$^{-3}$). We thus choose prior distributions that are motivated by the observed properties of long-duration GRBs. Log-normal priors are adopted for the microphysics parameters, centered on $\log_{10}\epsilon_e=-1$ \citep[$0.3$\,dex;][]{Beniamini2017} and $\log_{10}\epsilon_B=-2.5$ \citep[$1$\,dex;][]{Santana2014}. For the ISM density, we adopt $\log_{10}(n/\mathrm{cm^{-3}})\sim\mathcal{N}(0.0,\,1.45)$, a log-normal fitted to a compiled sample of long-GRB ISM densities \citep[$N=24$;][]{Panaitescu2002,Aksulu2022,Cenko2011}, whose $\sim1.5$\,dex scatter sets the prior width (\autoref{fig:density_prior}). We further place a log-normal prior on the jet half-opening angle centered on $\theta_c=5^\circ$ \citep[$0.35$\,dex, truncated to $0.2$--$20^\circ$;][]{Frail2001}, $p\sim\mathcal{N}(2.3,0.25)$ \citep{Curran2010}, and a log-uniform prior on $\Gamma_0$ over $[3,1000]$ \citep{Ghirlanda2018}. The viewing angle $\theta_{\rm obs}$ is left free with a uniform prior to allow for off-axis geometries. This information is summarized in \autoref{tab:priors}.

To test sensitivity to jet geometry, we run all models with 4000 live points. \naut\ provides a direct and robust comparison between models by returning both their posterior distributions and their Bayesian evidence ($\ln \mathcal{Z}$), from which a Bayes factor ($\ln \mathcal{B}$) can be calculated. As shown in \autoref{tab:evidence}, we find that the top-hat geometry is marginally preferred, though all jet geometries produce reliable fits to the \tess\ light curve. Therefore, the \tess\ data are unable to convincingly distinguish between the geometries, and so we simply select the top-hat as our fiducial model.

\begin{table}
\centering
\caption{Priors for the \event\ afterglow fit (\vegas\,+\,\naut; forward shock
only, lateral spreading disabled). Priors are denoted such that $\mathcal{U}$ is
linear-uniform, $\log_{10}\mathcal{U}$ is log-uniform, $\mathcal{N}$ is a normal
distribution, and $\log_{10}\mathcal{N}$ is log-normal. The microphysical and
geometric priors are motivated by the observed long-duration GRB distributions.}
%\citep{Curran2010,Beniamini2017,Santana2014,Frail2001,Ghirlanda2018,Wang2015,Aksulu2022}.}
\label{tab:priors}
\begin{tabular*}{\columnwidth}{@{\extracolsep{\fill}}lc@{}}
\hline\hline
Parameter & Prior \\
\hline
$\log_{10}(E_{\mathrm{K,iso}}/{\rm erg})$ & $\mathcal{U}(49,\,57)$ \\
$\log_{10}(n/{\rm cm^{-3}})$ & $\mathcal{N}(0,\,1.45)$ \\
$p$ & $\mathcal{N}(2.3,\,0.25)$ \\
$\Gamma_0$ & $\log_{10}\mathcal{U}(3,\,1000)$ \\
$\theta_{\rm c}$ & $\log_{10}\mathcal{N}(5^\circ,\,0.35\,{\rm dex})$ \\
$\theta_{\rm obs}$ & $\mathcal{U}(0,\,90)^\circ$ \\
$\log_{10}\epsilon_e$ & $\mathcal{N}(-1,\,0.3)$ \\
$\log_{10}\epsilon_B$ & $\mathcal{N}(-2.5,\,1.0)$ \\
$t_0$ (MJD$-59177$) & $\mathcal{U}(0.46,\,0.63)$ \\
$z$ & Joint \texttt{EAZY} $P(z)$ \\
\hline
\end{tabular*}
\end{table}

\begin{table}
\centering
\caption{Bayesian evidence comparison of jet geometries for \event\ modeled with
\vegas+\naut\ using 4000 live points. $\ln B$ is relative to the top-hat, which is
the marginally statistically preferred model. We are unable to differentiate
between jet geometries with just the \tess\ data.}
\label{tab:evidence}
\begin{tabular*}{\columnwidth}{@{\extracolsep{\fill}}lcccl@{}}
\hline\hline
Jet model & $\ln\mathcal{Z}$ & $\ln B$ & $\chi^2/{\rm dof}$ & Notes \\
\hline
Top-hat & $194.5$ & $0$ & $1.09$ & bimodal $\Gamma_0$ \\
Gaussian & $193.4$ & $-1.1$ & $1.08$ & $\Gamma_0\!\sim\!150$ \\
Power-law ($k{=}2$) & $192.2$ & $-2.3$ & $1.08$ & $\Gamma_0\!\sim\!80$ \\
\hline
\end{tabular*}
\end{table}

\begin{figure*}
    \includegraphics[width=\textwidth]{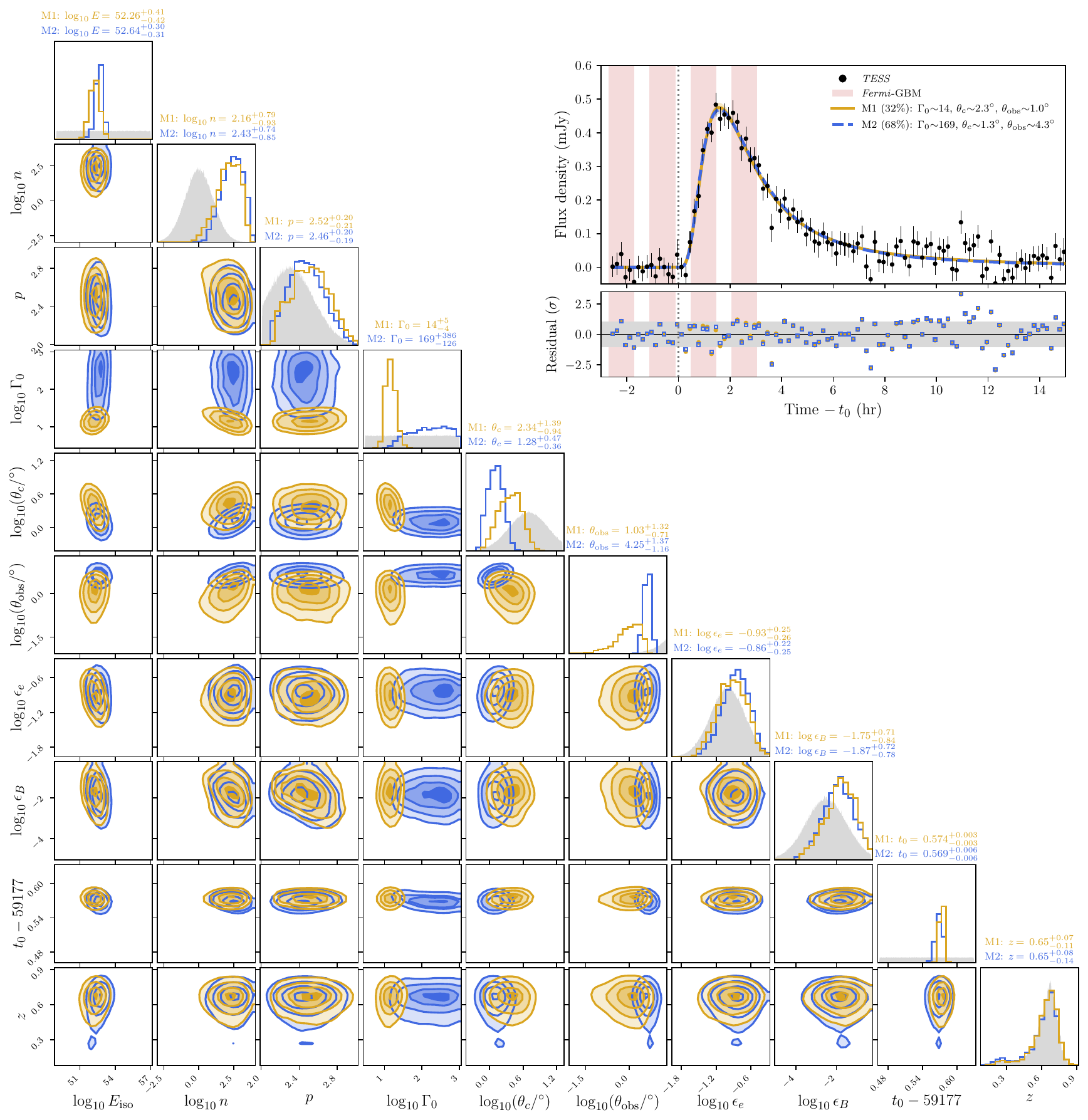}
    \caption{Combined plot for the on-axis \vegas\ fit to \event. \textbf{Left:} Corner plot of the \naut\ priors (gray) and posteriors, which are bimodal; a Gaussian-mixture decomposition splits them into Mode~1 (M1; gold, on-axis low-$\Gamma_0$) and Mode~2 (M2; blue, off-axis narrow jet with typical $\Gamma_0$), with each mode's median and $16$--$84$th percentiles above the diagonals. The two have equal evidence ($\Delta\ln\mathcal{Z}\simeq0$) and represent a $\Gamma_0$--geometry degeneracy, with the population fraction primarily driven by the broad posterior of M2. \textbf{Upper right:} The two mode-median models compared to the data (black points): M1 (gold, solid) and M2 (blue, dashed) with shaded $16$--$84\%$ regions and residuals below. The explosion time $t_0$ is marked, alongside red regions with \textit{Fermi}-GBM coverage near $t_0$.}
    \label{fig:vegasfit}
\end{figure*}

Within this geometry, we find clear bimodal structure in the viewing angle posterior distribution, corresponding to an on-axis and off-axis solution. We separate the posteriors into two distinct populations by fitting a two-component Gaussian mixture model to the standardized samples in the ($\theta_{\rm obs}$, $\log\Gamma_0$, $\log n$) subspace, assigning each sample to the on- or off-axis model by its maximum-probability component. As seen in \autoref{fig:vegasfit}, the two models share narrow jet opening angles ($\theta_c\approx1-2\rm^o$) and high ISM densities ($n\approx200$~cm$^{-3}$), but exhibit distinct characteristics: \textbf{Model 1} (subdominant 32\%) is an on-axis afterglow with a low initial bulk Lorentz factor ($\Gamma_0\approx14$); and \textbf{Model 2} (dominant 68\%) is an off-axis afterglow with a broad range of initial bulk Lorentz factors consistent with typical GRB afterglows. The values for all fit parameters for both models are shown in \autoref{tab:bimodal}. 

\begin{table}
\centering
\caption{The two-population posterior for the full-geometry top-hat fit of \event\
with \vegas+\naut\ and 4000 live points. We recover two solutions: an on-axis,
low-$\Gamma_0$ model (M1) and an off-axis, higher-$\Gamma_0$ model (M2). We
separate these solutions with a two-component Gaussian mixture in
$(\theta_{\rm obs},\Gamma_0,\log n)$. Both models fit the data equally well.
Values are posterior medians with $16$--$84$th-percentile uncertainties.}
\label{tab:bimodal}
\begin{tabular*}{\columnwidth}{@{\extracolsep{\fill}}lcc@{}}
\hline\hline
Parameter & M1 (on-axis) & M2 (off-axis) \\
\hline
Posterior fraction & 32\% & 68\% \\
\hline\\[-3mm]
$\log_{10}(E_{\rm iso}/{\rm erg})$ & $52.26^{+0.41}_{-0.42}$ & $52.64^{+0.30}_{-0.31}$ \\[1mm]
$\log_{10}(n/{\rm cm^{-3}})$ & $2.16^{+0.79}_{-0.93}$ & $2.43^{+0.74}_{-0.85}$ \\[1mm]
$p$ & $2.52^{+0.20}_{-0.21}$ & $2.46^{+0.20}_{-0.19}$ \\[1mm]
$\Gamma_0$ & $14^{+5}_{-4}$ & $169^{+386}_{-126}$ \\[1mm]
$\theta_c$ (deg) & $2.3^{+1.4}_{-0.9}$ & $1.3^{+0.5}_{-0.4}$ \\[1mm]
$\theta_{\rm obs}$ (deg) & $1.0^{+1.3}_{-0.7}$ & $4.3^{+1.4}_{-1.2}$ \\[1mm]
$\log_{10}\epsilon_e$ & $-0.93^{+0.25}_{-0.26}$ & $-0.86^{+0.22}_{-0.25}$ \\[1mm]
$\log_{10}\epsilon_B$ & $-1.75^{+0.71}_{-0.84}$ & $-1.87^{+0.72}_{-0.78}$ \\[1mm]
$t_0$ (MJD$-59177$) & $0.574^{+0.003}_{-0.003}$ & $0.569^{+0.006}_{-0.006}$ \\[1mm]
$z$ & $0.65^{+0.07}_{-0.11}$ & $0.65^{+0.08}_{-0.14}$ \\[1mm]
\hline
\end{tabular*}
\end{table}

We verify the stability of our solutions through parameter recovery simulations. For both models, we simulate 10 light curves with similar properties (SNR, cadence, etc.) to the observed data and rerun our fitting process. The recovery histograms are shown in \autoref{fig:recovery}. The solution for both models appears robust, with input parameters recovered within $2\sigma$ from this small sample size. Thus we infer that both high- and low-$\Gamma_0$ solutions are possible, and the \tess\ light curve alone is unable to conclusively distinguish between them.

Both Models 1 \& 2 provide excellent and indistinguishable fits to the data, returning well-behaved and stable posteriors. Notably, each suggest that \event\ arises from a gamma-quiet progenitor; an on-axis dirty fireball where the jet is choked by baryonic matter with $\Gamma_0\approx14$, or an off-axis orphan afterglow ($\theta_c< \theta_{obs}$). Furthermore, the models strongly prefer a dense ISM of $n >100\;\rm cm^{-3}$, a condition typically associated with molecular cloud environments. We discuss the implications of this model further in Subsection~\ref{sec:model_interp}.

\section{Discussion}\label{sec:dicsussion}
The \texttt{TESSELLATE} pipeline has unlocked the ability for \tess\ to lead discovery of fast extragalactic transients. Alongside a precision PSF light curve, \texttt{TESSELLATE} provides unmatched localization and photometric calibration that allow us to explore the nature of transients with just \tess\ data alone. In this study, we conclude that \event\ belongs in the family of GRB afterglows; in this section we evaluate our interpretation and its limitations.

\subsection{Model Interpretation} \label{sec:model_interp}

While \event\ occupies a similar location in duration-brightness parameter space to GRB afterglows, its light curve morphology most closely resembles those of the peculiar afterglows GRB~080710 and GRB~260310A. With \vegas\ and \naut, we conclude that \event\ is best described by either an off-axis orphan afterglow or an on-axis dirty fireball. Although we are unable to distinguish between these two models, each is interesting in its own right; \event\ is the first optically-discovered slow-evolving GRB afterglow with a precisely constrained explosion and peak time, alongside a well-resolved rise.

Confirming either classification hinges critically on the non-detection of prompt gamma-ray emission around $t_0$. For \event, this condition is impossible to satisfy; in 2020, GRB monitors such as \textit{Fermi}-GBM and \textit{Swift} had only patchy sky coverage. Through our modeling, we constrain the explosion time to $t_0=59177.574^{+0.003}_{-0.003}$~MJD for M1, and $t_0=59177.569^{+0.006}_{-0.006}$~MJD for M2, a window of just minutes. Unfortunately, there was no gamma-ray monitoring coverage of \event's position during this window; we indicate \textit{Fermi}'s temporal coverage of the position as pink shaded regions on \autoref{fig:vegasfit}. While there are no reported \textit{Fermi}-GBM detections during the lifetime of the event, the lack of coverage at $t_0$ prevents us from definitively classifying \event\ as the afterglow of a gamma-quiet GRB.

Another notable outcome from our fiducial model is the preference for high ISM density. While typical GRBs are expected to have low-density circumburst media, the population of long GRBs with density estimates, show a broad range extending into high densities consistent with molecular clouds ($n>100\;$cm$^{-3}$; see \autoref{fig:density_prior}). For \event, the densities for M1 and M2 are $n = 145^{+760}_{-128}\;{\rm cm^{-3}}$ and $n = 267^{+1214}_{-229};{\rm cm^{-3}}$, respectively, on the high end of the density distribution. As a consequence of this density, the synchrotron self-absorption frequency $\nu_a \propto n^{3/5}$ is shifted into the GHz range, effectively making the transient radio-quiet. Conversely, IR wavelengths are boosted in intensity relative to optical, as the cooling frequency ($\nu_{c}\propto n^{-1}$) is pushed into the optical regime at the densities returned by the model.

Both models point to a collapsar as the likely origin of \event. For Model 1 -- a moderately relativistic blast decelerating in a dense medium -- the collapsar launches a baryon-loaded, low-$\Gamma_0$ jet that powers a luminous synchrotron afterglow while its prompt emission is stifled by pair-production opacity \citep{Huang2002}. Conversely, Model 2 -- an off-axis orphan afterglow -- aligns with the model proposed for GRB~260310A which was associated with the Type Ic supernova  SN~2026fgk. Alternative progenitor explanations are disfavored by our modeling. The dense medium rules out a compact-binary merger, since these occur in low-density environments \citep{Berger2014}.

If \event\ was indeed produced by a collapsar in a dense molecular environment, then evidence for this origin may still be observable with \textit{JWST}. A collapsar would leave a lasting mid-infrared signature from dust, via two channels: pre-existing circumburst grains reprocessing the luminous transient into a thermal echo, such as recently seen in the stripped-envelope SN~2024aecx \citep{Tinyanont2026}; and newly formed dust in the expanding ejecta, which is now routinely revealed by \textit{JWST} in core-collapse events, including Type~Ic and broad-lined Ic supernovae \citep{Shahbandeh2023,Ravi2023,Shrestha2026}. At the current epoch ($\sim3.4$ rest-frame years post-burst), the echo would arise from dust at $R\sim0.5$~pc, heated by the transient to $\sim10^3$~K and radiating at rest-frame $\sim1.5$--$5\,\mu$m; at $z=0.67$ this shifts to an observed $\sim2.5$--$8\,\mu$m. Under plausible assumptions the echo luminosity may reach $\sim27$~AB making it visible to \textit{JWST}'s NIRCam/MIRI. Therefore, targeted \textit{JWST} observations of \event's localization could provide an independent confirmation on the collapsar origin of \event\ years after explosion.

\subsection{Host Identification}

In this analysis, we have assumed that \event\ is associated with the galaxies we identified as components A and B. Our analysis of available photometry suggests that these may be interacting, due to the similar \texttt{EAZY} $z_{\rm phot}$ distribution, brightness, and proximity. Furthermore, the SEDs modeled by \texttt{prospector} are consistent with low-mass star-forming galaxies, in line with expectations of an interacting galaxy pair. These results align with our collapsar interpretation of \event, as such progenitors predominantly occur in star forming regions of galaxies \citep{Fruchter2006,Blanchard2016}. This supports the compelling narrative for the origins of \event\ that arise from modeling.

A key issue facing this analysis is the event localization. While \tess\ has large $21''$ pixels, our PSF-driven localization procedure confidently recovers the position of \event\ to $\sim1''$, but returns an offset of $2.6''^{+1.2}_{-0.7}$ from the center of galaxy A. As outlined previously, at a redshift of $\sim0.67$, this on-sky offset corresponds to a physical separation of $18.9^{+8.6}_{-5.1}$ kpc, placing \event\ $2.7\sigma$ from the core of galaxy A, presenting a mild tension. Crucially, dense star forming regions required for the collapsar interpretation of \event\ are unlikely to exist at such a separation; however, it is possible that a faint tidal knot is present closer to the source position. Alternatively, the true host may be a faint galaxy below DECam detection limits. Further spectroscopic and deep imaging observations of the region are needed to resolve the host uncertainty.

\subsection{Gamma-ray Quiet Afterglows}

While gamma-ray quiet afterglows, such as dirty fireballs and off-axis orphans, were predicted to be ubiquitous \citep{Huang2002}, they have proven an elusive class of transient. In recent years, wide-field optical surveys have detected several relativistic transients unaccompanied by a GRB trigger, such as PTF\,11agg \citep{Cenko2013}, AT~2020blt \citep{Ho2020}, and AT~2019pim \citep{Perley2025}. However, identifying true gamma-quiet events remains challenging due to patchy coverage of monitors like Fermi-GBM. Furthermore, distinguishing between an on-axis dirty fireball and an off-axis orphan is challenging, and not possible with \tess\ data alone. 

The challenge persists even When multiple bands are available. Even with radio observations it is challenging to classify AT~2019pim, the first dirty fireball candidate observed by \tess. Unlike \event, AT~20019pim was observed at a slower \tess\ cadence of 30~minutes, which does not reliably capture the explosion time. A reanalysis of AT~2019pim with our framework indicates inconsistencies in the reported \textit{Fermi}-GBM timings, and strongly favors a structured jet with $\Gamma_0 >100$. 
Our results, which suggest that AT~2019pim may be an off-axis afterglow instead of a dirty fireball will be discussed in upcoming work.

Our 10-minute-cadence light curve for \event\ potentially places it as the first optically-discovered gamma-quiet afterglow with time-resolved evolution. It shares, however, the same fundamental limitation as its predecessors: \textit{Fermi}-GBM was Earth-occulted at $t_0$, and thus we cannot establish whether the prompt emission was intrinsically suppressed or merely unobserved. \event\ therefore joins the handful of optically-detected relativistic transients that populate a new extension of the GRB population.

\subsection{Future Prospects}

\event\ was detected as a high-SNR ($>10\sigma$) event in the pilot high-galactic latitude sub-survey from our \tessellate\ program, which is analyzing all year 3 and 4 data from \tess. The discovery of \event\ proves that \tess\ is capable of discovering rare classes of fast extragalactic transients, establishing it as a discovery engine in its own right, rather than a follow-up resource alone. We anticipate that many more extragalactic transients have been detected and documented at lower SNR by \tessellate\ in this pilot survey; however, at these lower significances, instrumental artifacts and other astrophysical contaminants become far more common, requiring sophisticated filtering methods to isolate high-quality extra-galactic candidates. We are currently developing frameworks to more effectively automate the classification and filtering of events discovered by \tessellate.

While it is likely too late to obtain targeted late-time follow-up for \event, \tess\ continues to observe the sky at a near continuous cadence. Following the completion of the pilot survey programs, \tessellate\ will be run on newly released \tess\ sectors, enabling deep optical and radio follow-up of unique fast extragalactic transients to be triggered within days to weeks of discovery. Combining the current 200s cadence \tess\ data with such observations will allow us to develop clearer interpretations for any fast transients we discover.

\section{Conclusion}
In this paper we present the first fast extragalactic transient discovered by \tess, made possible with the \tessellate\ pipeline. \event\ was identified through the pilot program of our high galactic latitude sub-survey (HiLaTS) which has analyzed FFIs from years 3 and 4 of \tess\ operations --- the first years of 10 minute cadence imaging. We localized \event\ to be offset from galaxy J042144.37-383311.3 by $2.6''^{+1.2}_{-0.7}$. 
Photometric modeling suggests that this galaxy is a low-mass star-forming galaxy that is interacting with a larger companion J042144.37-383313.0 at a redshift of $z_{\rm phot}= 0.67^{+0.07}_{-0.10}$. 

The fast evolving and luminous nature of \event\ suggests its progenitor is a relativistic engine, most plausibly manifesting as a GRB afterglow. With a robust modeling pipeline that uses the \vegas\ and \naut\ packages, we find \event\ is best described either by an off-axis orphan afterglow or an on-axis, low $\Gamma_0$ dirty fireball. Despite its high SNR and temporal resolution, which unambiguously resolves the evolution of \event, the \tess\ light curve is unable to disentangle these two models, highlighting the difficulty of robust physical parameter estimation from single-band photometry alone. Regardless, \event\ stands as a benchmark example of either class with both a well-constrained explosion and peak time.

Our interpretations of \event\ are limited in two key ways. The first is the lack of coverage from \textit{Fermi}-GBM or \textit{Swift}-BAT at the time of explosion, which prevents confirmation of the absence of a prompt gamma-ray precursor, a key requirement for classifying either an off-axis or dirty fireball afterglow. The second is host association; while \event\ appears coincident with galaxy J042144.37-383311.3 (A), the localization places it $18.9^{+8.6}_{-5.1}$ kpc from the galaxy's center --- beyond the typical separation distance for long GRBs. Deep observations of the field are needed to identify if there is a tidal knot from the interacting galaxies, or a faint galaxy closer to the event centroid that exists below the DECam detection limit. 

Regardless, the discovery of \event\ marks a new era for \tess\ in transient analysis. We have demonstrated its capability as a tool for leading the discovery of fast extragalactic transients, rather than simply a follow-up resource. Its unique high-cadence, near-continuous observation strategy allows us to temporally resolve rapid transients, which is crucial to understanding their nature and the physical mechanisms from which they arise. While follow-up observations are challenging for this six-year-old event, \tessellate\ will soon be applied to newly released \tess\ sectors, enabling deep multi-wavelength follow-up to be triggered for fast transients within weeks or days of discovery. As such, \tess\ is poised to further solidify its strength in the optical time-domain space, and potentially uncover a population of fast extragalactic transients that has, until now, remained largely hidden from view.

\begin{acknowledgments}
R.R-H., Z.G.L., and C.M. are supported by the Royal Society of New Zealand, Te Ap\={a}rangi through the Marsden Fund Fast Start Grant M1255, and M.T.B by by the Rutherford Discovery Fellowships from New Zealand Government funding. 

H.R. and B.L. are supported by Australian Government Research Training Program (RTP) Scholarships.

This work was partially supported by NASA through award number 80GSFC24M0006 and the ADAP grant 80NSSC22K0494.

Part of this research was funded by the Australian Research Council CE230100016. J.C. acknowledges funding by the Australian Research Council Discovery Project, DP200102102

This paper includes data collected with the \tess\ mission, obtained from the MAST data archive at the Space Telescope Science Institute (STScI). Funding for the TESS mission is provided by the NASA Explorer Program. STScI is operated by the Association of Universities for Research in Astronomy, Inc., under NASA contract NAS 5–26555.

This work was performed on the OzSTAR national facility at Swinburne University of Technology.
The OzSTAR program receives funding in part from the Astronomy National Collaborative Research Infrastructure Strategy (NCRIS) allocation provided by the Australian Government, and from the Victorian Higher Education State Investment Fund (VHESIF) provided by the Victorian Government.

\end{acknowledgments}

\vspace{5mm}
\facilities{\tess, OzSTAR, 
DECam, WISE, \textit{Gaia}, MAST
}

\section*{Data Availability}

This paper is based on \tess\ full-frame images, which are publicly available from the Mikulski Archive for Space Telescopes (MAST) at the Space Telescope Science Institute\footnote{\url{https://mast.stsci.edu}}. \event\ was observed by \tess\ in Sector~32 (Camera~3, CCD~2); the calibrated full-frame images are released through the MAST archive \dataset[doi:10.17909/0cp4-2j79]{https://doi.org/10.17909/0cp4-2j79}. The reduced \tess\ light curve of \event\ used in this work is provided in \autoref{tab:lc}. The archival imaging used for the host association and photometric-redshift analysis is likewise public: DECam imaging from the DESI Legacy Imaging Surveys (DR10), \textit{WISE}/unWISE, and \textit{Gaia}.

The \tessellate\ transient-detection pipeline \citep{TESSELLATE} and the \texttt{TESSreduce} reduction package \citep{tessreduce} are open source and available at \url{https://github.com/rhoxu/TESSELLATE} and \url{https://github.com/CheerfulUser/TESSreduce}, respectively. The afterglow modeling was performed with the public codes \vegas\ \citep{Wang2026} and \naut\ \citep{nautilus}. The \vegas\ model configurations and the \naut\ posterior samples underlying the figures and tables in this paper are available at \url{https://github.com/CheerfulUser/at2020afjz}.

\software{
\tessellate\ \citep{TESSELLATE},
% \ttt{SourceDetect} \citep{Moore2025},
\texttt{TESSreduce} \citep{tessreduce},
\texttt{astropy} \citepalias{Astropy2013,Astropy2018,Astropy2022},
\texttt{astrocut} \citep{astrocut},
\texttt{astroquery} \citep{astroquery},
\texttt{EAZY} \citep{Brammer2008},
\texttt{emcee} \citep{Foreman-Mackey2013},
\naut\ \citep{nautilus},
\texttt{numpy} \citep{numpy},
\texttt{matplotlib} \citep{Hunter2007},
\texttt{Prospector} \citep{Johnson2021},
\texttt{pandas} \citep{mckinney-proc-scipy-2010,reback2020pandas},
\texttt{photutils} \citep{Bradley2024},
\texttt{scikit-learn} \citep{scikitlearn2011},
\texttt{scipy} \citep{2020SciPy-NMeth},
\texttt{TESS\_PRF} \citep{tessprf},
\vegas\ \citep{Wang2026},
\texttt{corner} \citep{corner}.
}

% \clearpage

\bibliography{main}{}
\bibliographystyle{aasjournalv7}


\clearpage
\appendix
\restartappendixnumbering
\section{Localization}
\begin{figure*}[!h]
    \includegraphics[width=1\textwidth]{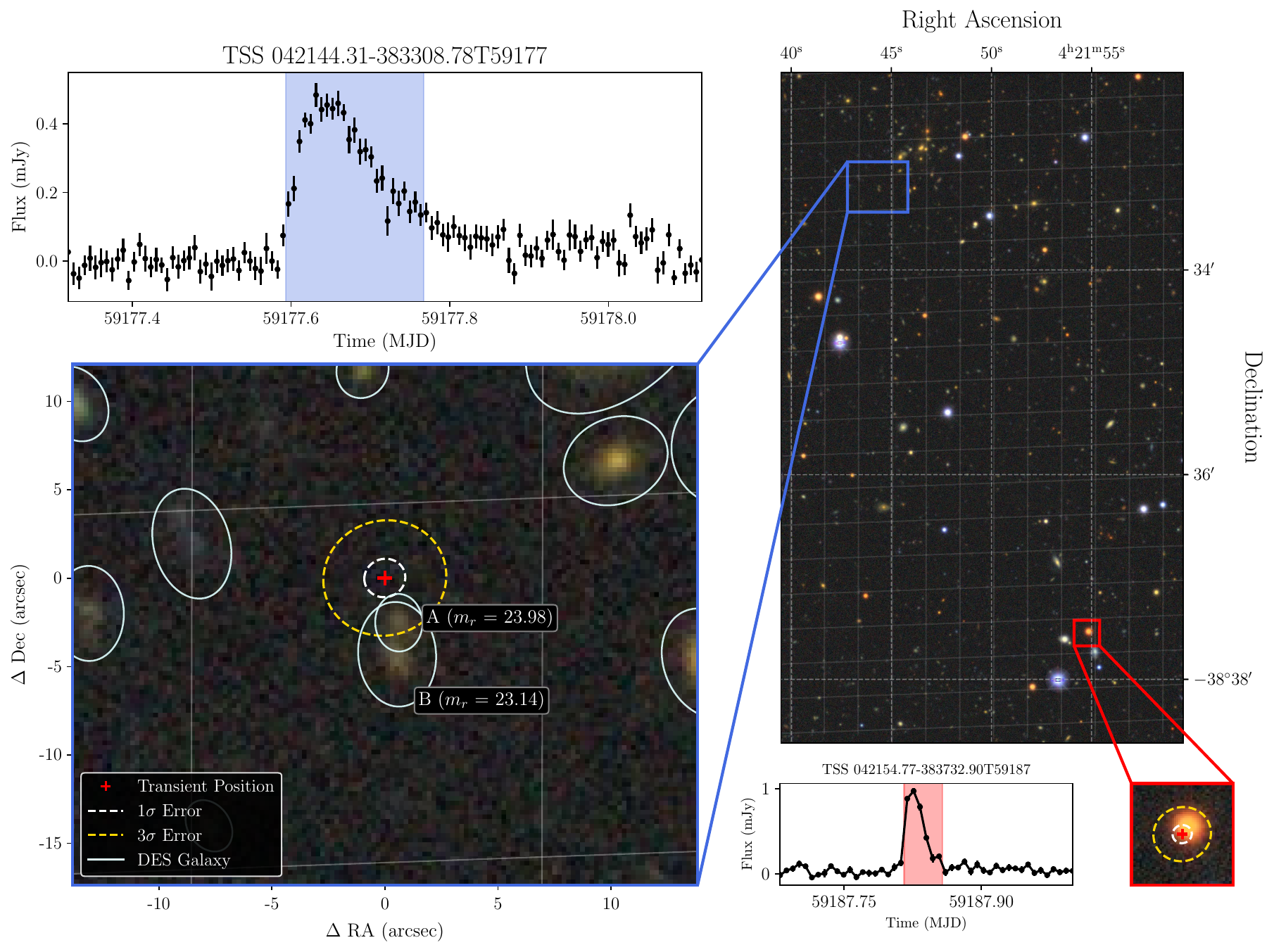}
    \caption{Verification of World Coordinate System quality in the local \tess\ field around \event. The right panel shows Legacy DR10 DECam imaging for a wider region near the localization box in blue, with \tess\ pixels indicated by the faint grid lines. In the bottom right corner, $\sim$ 14 pixels from \event, a flare star experienced an outburst which was detected by \texttt{TESSELLATE}; the light curve and localization of this event are shown in the bottom right panels. The accuracy of the flare star localization is very high, confirming the quality of \event's localization.}
    \label{fig:complex}
\end{figure*}

One of the key challenges in identifying likely extragalactic transients observed by \tess\ is in localizing the position, given the 21'' pixels. With \tessellate\ we have created a robust method for localizing sources to within $\sim10\%$ of the pixel size. In \autoref{fig:complex} we show the positive association of a stellar flare with a star. The stellar flare was just 14 \tess\ pixels away from the position of \event\ confirming the accuracy of our world coordinate system for transient events.

\section{Host properties}
\begin{figure*}
    \includegraphics[width=\textwidth]{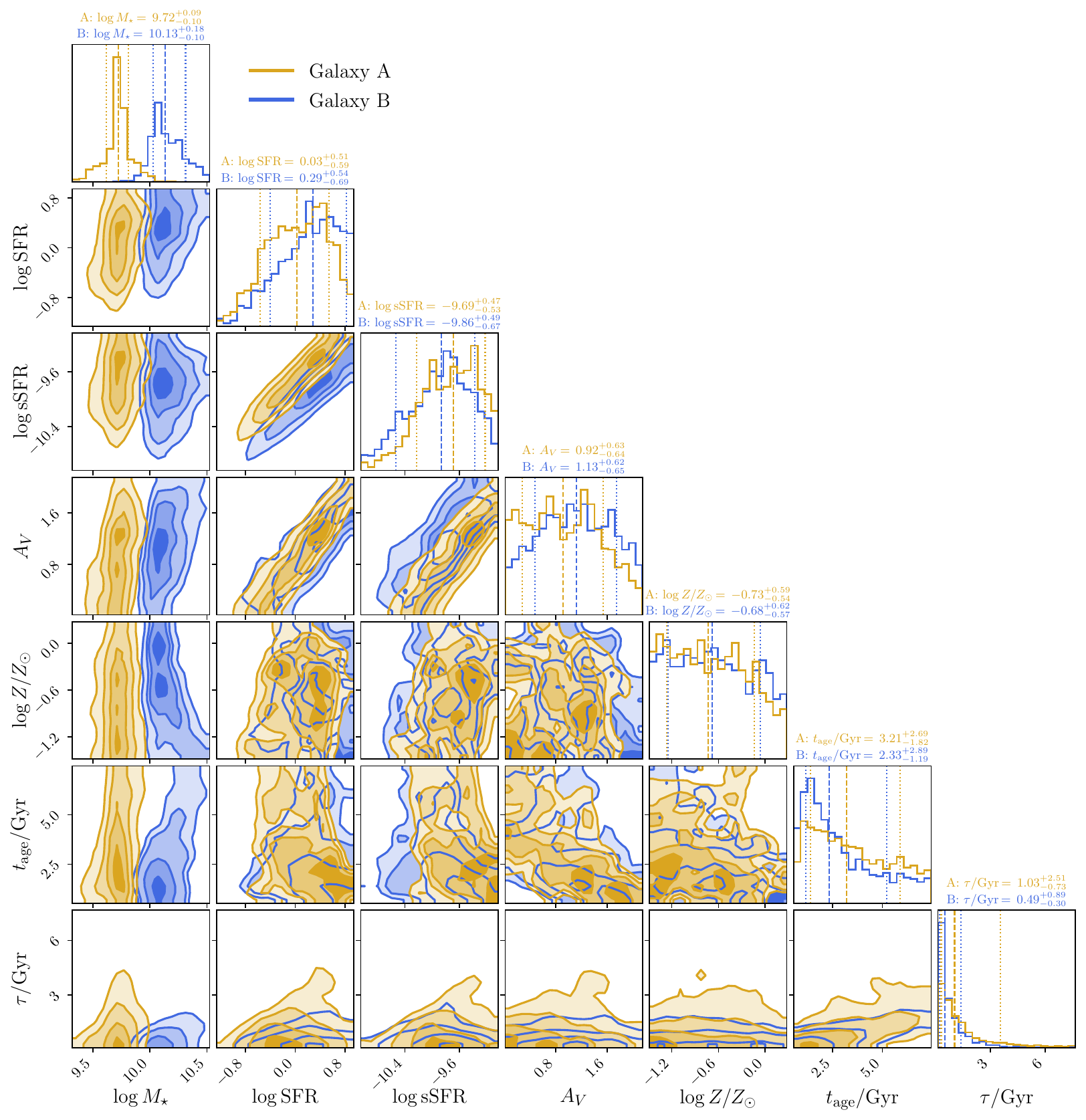}
    \caption{Corner plot of the \texttt{Prospector} posterior distributions for galaxies A (gold) and B (blue) assuming a redshift of 0.67. Although the available data is limited, the masses is well constrained,  while the low e-folding timescale of the star-formation history suggests there is ongoing star formation. The other parameters are prior dominated.}
    \label{fig:prospect_corner}
\end{figure*}

We associate \event\ to the faint galaxies DES J042144.37-383311.3 and DES J042144.37-383313.0 (A and B). With the available photometry from DECam and unWISE, we constrain some physical properties of the galaxies. In \autoref{fig:prospect_corner} we show the posterior corner plot from the \texttt{Prospector} fit to the photometry. 

\section{Model fitting}

\begin{figure}
    \centering
    \includegraphics[width=0.5\columnwidth]{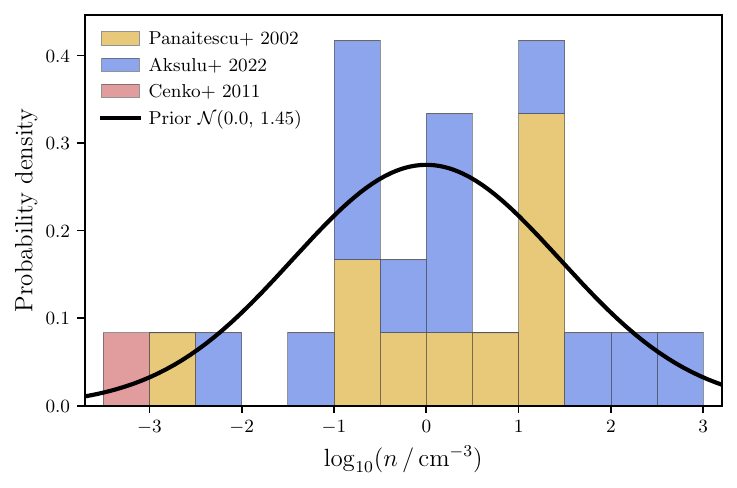}
    \caption{Circumburst density distribution of long-duration GRBs modeled with a homogeneous (ISM) medium, compiled from broadband afterglow fits in the literature: \citet[][gold]{Panaitescu2002}, \citet[][blue]{Aksulu2022}, and \citet[][red]{Cenko2011}, for a total of $N=24$ measurements spanning $\log_{10}(n/{\rm cm^{-3}})$ from $-3.2$ to $2.6$. The black curve is a Gaussian fit to the pooled sample in $\log_{10}n$ ($\mu=0.02$, $\sigma=1.46$), which we adopt (rounded to $\mathcal{N}(0.0,\,1.45)$) as the density prior for our \event\ afterglow fit.}
    \label{fig:density_prior}
\end{figure}

Through preliminary testing we found that \vegas\ fits to \event\ were driven to unphysical ISM densities of $n>10^5\,$cm$^{-3}$ when using broad flat priors. Given the degeneracy between the angles, $\Gamma_0$, and $n$, we construct an informed prior for $n$ by fitting a simple density function to reported densities for long GRBs. The ISM density measurements, and the prior model are shown in \autoref{fig:density_prior}.

\begin{figure}
    \includegraphics[width=1\textwidth]{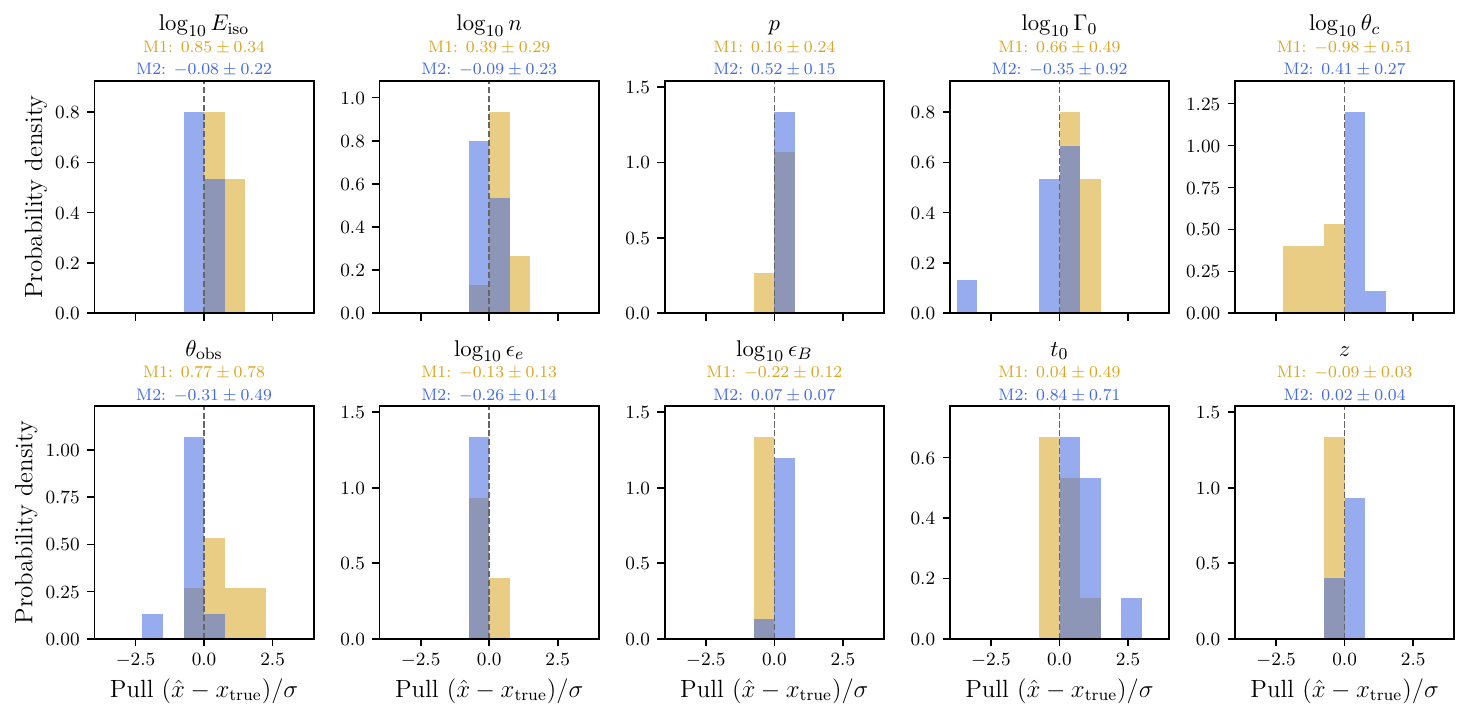}
    \caption{Injection--recovery pull distributions for the two top-hat solutions, M1 (on-axis, gold) and M2 (off-axis, blue). For each mode we simulate $N=10$ \tess\ light curves from that mode's median parameters --- matched to the observed cadence, SNR, and errors --- and refit them with the identical \vegas+\naut\ pipeline. The pull is $(\hat{x}-x_{\rm true})/\sigma$; each panel is labeled with the parameter and the per-mode mean and width of its pull distribution, and an unbiased, well-calibrated recovery yields mean $\simeq0$ and width $\simeq1$ (dashed line marks zero). Both modes recover their input parameters within $\sim2\sigma$ for this small sample, with widths $\lesssim1$ indicating conservative uncertainties. Neither solution is preferred, confirming that the \tess\ light curve alone cannot distinguish the on- and off-axis interpretations.}
    \label{fig:recovery}
\end{figure}

Since afterglow models can be highly degenerate, it is essential to test parameter retrieval. We simulate 10 light curves of similar quality to \event\ using the fiducial model and perform the same fitting procedure and compare the recovered parameters to the input values. In \autoref{fig:recovery} we show the pull distributions for each of the fit parameters for both models.

\newpage
\section{Flux table}
We provide the light curve of \event\ for the time interval from 1.5 hours before peak to 6.3 hours after peak in \autoref{tab:lc}. 

\startlongtable
\begin{deluxetable}{ccccc}
\tablecaption{\textit{TESS} photometry of \event\ from 1.5 hours before peak to 6.3 hours after peak. Flux density and magnitudes are on the AB system (zero point $20.626$).
Magnitudes are reported as $m\pm1\sigma$ for $\geq3\sigma$ detections; all other
epochs are given as $3\sigma$ upper limits (denoted ``$>$'').
Absolute magnitudes assume the host photometric redshift $z=0.67$ and a
$k$-correction $2.5(1-\beta)\log_{10}(1+z)$ with $\beta=(p-1)/2=0.65$
($M=m-\mathrm{DM}(z)+k$, $\mathrm{DM}=43.09$).\label{tab:lc}}
\tablehead{\colhead{Time} & \colhead{$F_\nu$} & \colhead{$m_{\rm AB}$} & \colhead{$M_{\rm AB}$} \\
\colhead{(MJD)} & \colhead{(mJy)} & \colhead{(mag)} & \colhead{(mag)}}
\startdata
59177.57607 & $0.0003\pm0.0287$ & $>19.06$ & $>-23.84$ \\
59177.58301 & $-0.0234\pm0.0266$ & $>19.15$ & $>-23.75$ \\
59177.58996 & $0.0749\pm0.0313$ & $>18.97$ & $>-23.93$ \\
59177.59690 & $0.1667\pm0.0372$ & $18.35\pm0.24$ & $-24.55\pm0.24$ \\
59177.60384 & $0.2115\pm0.0364$ & $18.09\pm0.19$ & $-24.81\pm0.19$ \\
59177.61079 & $0.3482\pm0.0326$ & $17.55\pm0.10$ & $-25.35\pm0.10$ \\
59177.61773 & $0.4109\pm0.0221$ & $17.37\pm0.06$ & $-25.53\pm0.06$ \\
59177.62468 & $0.4003\pm0.0283$ & $17.39\pm0.08$ & $-25.50\pm0.08$ \\
59177.63162 & $0.4835\pm0.0349$ & $17.19\pm0.08$ & $-25.71\pm0.08$ \\
59177.63857 & $0.4412\pm0.0358$ & $17.29\pm0.09$ & $-25.61\pm0.09$ \\
59177.64551 & $0.4542\pm0.0336$ & $17.26\pm0.08$ & $-25.64\pm0.08$ \\
59177.65245 & $0.4440\pm0.0323$ & $17.28\pm0.08$ & $-25.62\pm0.08$ \\
59177.65940 & $0.4591\pm0.0375$ & $17.25\pm0.09$ & $-25.65\pm0.09$ \\
59177.66634 & $0.4321\pm0.0248$ & $17.31\pm0.06$ & $-25.59\pm0.06$ \\
59177.67329 & $0.3540\pm0.0392$ & $17.53\pm0.12$ & $-25.37\pm0.12$ \\
59177.68023 & $0.3824\pm0.0356$ & $17.44\pm0.10$ & $-25.46\pm0.10$ \\
59177.68718 & $0.3188\pm0.0371$ & $17.64\pm0.13$ & $-25.26\pm0.13$ \\
59177.69412 & $0.3246\pm0.0322$ & $17.62\pm0.11$ & $-25.28\pm0.11$ \\
59177.70107 & $0.3035\pm0.0299$ & $17.69\pm0.11$ & $-25.20\pm0.11$ \\
59177.70801 & $0.2337\pm0.0346$ & $17.98\pm0.16$ & $-24.92\pm0.16$ \\
59177.71495 & $0.2418\pm0.0363$ & $17.94\pm0.16$ & $-24.96\pm0.16$ \\
59177.72190 & $0.1169\pm0.0426$ & $>18.63$ & $>-24.26$ \\
59177.72884 & $0.2035\pm0.0378$ & $18.13\pm0.20$ & $-24.77\pm0.20$ \\
59177.73579 & $0.1679\pm0.0369$ & $18.34\pm0.24$ & $-24.56\pm0.24$ \\
59177.74273 & $0.2041\pm0.0279$ & $18.13\pm0.15$ & $-24.77\pm0.15$ \\
59177.74968 & $0.1446\pm0.0327$ & $18.50\pm0.25$ & $-24.40\pm0.25$ \\
59177.75662 & $0.1718\pm0.0313$ & $18.31\pm0.20$ & $-24.59\pm0.20$ \\
59177.76357 & $0.1345\pm0.0324$ & $18.58\pm0.26$ & $-24.32\pm0.26$ \\
59177.77051 & $0.1414\pm0.0285$ & $18.52\pm0.22$ & $-24.37\pm0.22$ \\
59177.77745 & $0.0971\pm0.0350$ & $>18.85$ & $>-24.05$ \\
59177.78440 & $0.1127\pm0.0341$ & $18.77\pm0.33$ & $-24.13\pm0.33$ \\
59177.79134 & $0.0762\pm0.0326$ & $>18.92$ & $>-23.97$ \\
59177.79829 & $0.0705\pm0.0427$ & $>18.63$ & $>-24.27$ \\
59177.80523 & $0.1014\pm0.0320$ & $18.88\pm0.34$ & $-24.01\pm0.34$ \\
59177.81218 & $0.0744\pm0.0266$ & $>19.14$ & $>-23.76$ \\
59177.81912 & $0.0683\pm0.0387$ & $>18.74$ & $>-24.16$ \\
59177.82607 & $0.0412\pm0.0369$ & $>18.79$ & $>-24.11$ \\
59177.83301 & $0.0723\pm0.0353$ & $>18.84$ & $>-24.06$ \\
59177.83995 & $0.0679\pm0.0338$ & $>18.88$ & $>-24.01$ \\
59177.84690 & $0.0647\pm0.0369$ & $>18.79$ & $>-24.11$ \\
59177.85384 & $0.0475\pm0.0331$ & $>18.91$ & $>-23.99$ \\
59177.86079 & $0.0706\pm0.0341$ & $>18.87$ & $>-24.03$ \\
59177.86773 & $0.0921\pm0.0314$ & $>18.97$ & $>-23.93$ \\
59177.87468 & $0.0023\pm0.0364$ & $>18.80$ & $>-24.10$ \\
59177.88162 & $-0.0355\pm0.0306$ & $>18.99$ & $>-23.90$ \\
59177.88856 & $0.0756\pm0.0336$ & $>18.89$ & $>-24.01$ \\
59177.89551 & $0.0178\pm0.0311$ & $>18.98$ & $>-23.92$ \\
\enddata
\end{deluxetable}

\end{document}